\documentclass[12pt]{article}
\usepackage{amssymb}
\usepackage{graphicx,color,epstopdf}
\usepackage{hyperref}
\def\hDash{\bot\!\!\!\bot}

\usepackage{amsfonts}
\usepackage{mathrsfs}
\usepackage{amsthm}
\usepackage{multirow}
\usepackage{bbding}
\usepackage{amssymb}
\usepackage{amsmath}
\usepackage{graphicx,color}
\usepackage{caption}
\usepackage{subcaption}
\usepackage{enumerate}
\usepackage{comment}
\usepackage{array}
\usepackage{bm}
\usepackage{booktabs}
\usepackage{titlesec}
\usepackage{dcolumn}
\usepackage[authoryear,round,]{natbib}

\newtheorem{theorem}{Theorem}[section]

\newtheorem{lemma}{Lemma}[section]
\newtheorem{remark}{Remark}[section]
\numberwithin{equation}{section}

\begin{document}

\title{An adaptive-to-model test for partially parametric single-index models
\footnote{Lixing Zhu is a Chair professor of Department of Mathematics
at Hong Kong Baptist University, Hong Kong, China. He was supported by a grant from the
University Grants Council of Hong Kong, Hong Kong, China. }}
\author{Xuehu Zhu$^{1}$,  Xu Guo$^{3}$ and Lixing Zhu$^1$\\
{\small {\small {\it$^1$ Hong Kong Baptist University, Hong Kong}}}\\
{\small {\small {\it $^3$ Nanjing University of Aeronautics and Astronautics, Nanjing } }}\\
}
\date{}
\maketitle

\renewcommand\baselinestretch{1.5}
{\small}

\noindent {\bf Abstract.} Residual marked empirical process-based tests are commonly used in regression models. However, they suffer from data sparseness in high-dimensional space when there are many covariates. This paper has three purposes. First, for partially parametric single-index models, we suggest a partial dimension reduction adaptive-to-model testing procedure to extend an existing directional test
into an omnibus test.  The resulting test is omnibus against general global alternative models. The procedure can automatically adapt to  the null and alternative models  to fully utilise the dimension-reduction structure under the null hypothesis, and thus greatly overcome the dimensionality problem. Second, to achieve the above goal, we propose a ridge-type eigenvalue ratio estimate to automatically determine the number of linear combinations of the covariates under the null and alternatives.  Third, a Monte-Carlo approximation to the sampling null distribution is suggested. Unlike existing bootstrap approximation methods, this gives an  approximation as close to the sampling null distribution as possible  by fully utilising the dimension reduction model structure under the null.  Simulation studies and real data analysis are then conducted  to illustrate the performance of the new test and compare it with existing tests.

\newpage
\baselineskip=21pt

\newpage

\setcounter{equation}{0}
\section{Introduction}
Consider the  partially parametric single-index model in the form:
\begin{equation}\label{(1.1)}
Y = G(\beta^{\top}X, W, \theta) + \epsilon,
\end{equation}
where $Y$ is the  response variable, $(X, W)$ is the covariate vector  in $\mathbb{R}^{p_1+p_2} $, $G(\cdot)$ is a known smooth function that depends not only on the covariate $\beta^{\top}X$ but also on the covariate $W$, $\beta$ and $\theta$ are the unknown regression parameter vectors and the error $\epsilon$ follows a continuous distribution and is independent with the covariates $(X,W)$. The model~(\ref{(1.1)})  reduces to the parametric single-index model in the absence
of the covariate $W$ and to the general parametric model in the absence of the covariate $\beta^{\top}X$. This structure is often meaningful, as in many applications, $p_1$ is large while $p_2$ is not. See the relevant dimension reduction literature, such as Feng et al. (2013). 

However, it is less clear whether a real data set fits the above statistical formalisation. It is worthwhile performing suitable and efficient model checking before any further statistical analysis. As we often have no idea about the model structure under the alternative hypothesis, the general alternative model is considered in the following form:
\begin{equation}\label{(1.2)}
Y = g(X, W) + \epsilon,
\end{equation}
where $g(\cdot)$ donates an unknown smooth function.

Several methods  for testing the parametric single-index model that removes  the covariate $W$ from the model~(\ref{(1.1)}), and the general nonlinear model in the absence of the covariate $\beta^{\top}X$can be found  in the literature. Two prevalent classes of method are local and global smoothing tests. A local smoothing test involves a nonparametric smoothing technique in the estimation, whereas a global smoothing test only requires a set of sample averages with respect to an index set to form an empirical process.  For examples, 
H\"{a}rdle and Mammen (1993)
suggested a local smoothing test based on the $L_2$ distance between the parametric and nonparametric estimate of the conditional expectation of $Y$ given $(X, W)$ in our notation.
Zheng (1996) and Fan and Li (1996)
independently developed tests based on  second order conditional moments. Dette (1999) proposed a consistent test that depended on the difference between the  variance estimate under the null and alternative hypotheses. Fan et al. (2001) developed a generalised likelihood ratio test. For other developments, see the Neyman threshold test (Fan and Huang 2001), a class of minimum distance tests (Koul and Ni 2004) and the distribution distance test (Van Keilegom et al. 2008). Gonz\'{a}lez-Manteiga and Crujeiras (2013) is a comprehensive review.
However, local smoothing tests have two obvious shortcomings. First, those methodologies have the subjective constraint choice of tuning parameters such as bandwidth.
Unlike estimation, finding an optimal bandwidth choice for hypothesis testing is still an open problem (Stute and Zhu 2005). Although practical evidence suggests that this issue is not critical when the number  $p$ of covariates  is small, a proper choice is not easy at all when $p$ is large, even moderate. This problem often results in poor  type I error. A more serious problem is  the typical slow convergence rates of local smoothing tests,  that is $O(n^{-1/2}h^{-p/4})$ under the null hypothesis, where $h$ is the bandwidth tending to zero. In the present setup, $p=p_1+p_2.$ In other words,   local smoothing tests suffer  severely from the curse of dimensionality.

For global smoothing tests, examples include Bierens (1990),
Stute (1997),  and  Khmaladze and Koul (2004).
Stute et al. (1998) used bootstrap approximation to determine the critical values of the residual-marked empirical process-based test. Resampling approximation is particularly required when $p$ is larger than $2$ as its limiting null distribution is intractable.
Escanciano (2007) is also a relevant reference in this class of tests.
The typical convergence rate of global smoothing tests is $O(n^{-1/2})$. Thus, they have the theoretical advantages over local smoothing tests.
However, 
the data sparseness
in high-dimensional space means that  most global smoothing tests suffer from the dimensionality problem, even for large sample sizes (see Escanciano, 2006).
Practical evidence  shows that  the power of global smoothing tests deteriorates and maintaining  the significance level becomes more difficult when the dimension $p$ of $X$ is large, or even moderate. This is particularly the case when the alternative model is high-frequency.

A direct way to  alleviate this problem is to project the high-dimensional covariates onto one-dimensional spaces first, and a test can be an average of tests that are based on the projections. This is a typical method called the projection-pursuit. Huber (1985) is a comprehensive reference.  Zhu and Li (1998) suggested using the projection pursuit technique to define a test based on   an unweighted integral of expectations with respect to all one-dimensional directions. Zhu and An (1992) has already used this idea to deal with a relevant testing problem.
Lavergne and Patilea (2008)  adopted this idea and further developed a dimension-reduction nonparametric method by exploring an optimal direction.
Lavergne and Patilea (2012) advised a smooth version of the integrated conditional moment test over all projection directions.
All of these tests partly overcome the curse of dimensionality with the use of one-dimensional projections. However, the computational burden is a serious issue. Computing the values of the test statistics is very time-consuming, and becomes even more serious if we further need to use bootstrap approximation to determine critical values. Based on our very limited numerical studies, which we do not report in this paper, the CPU consumption of such tests is more than 100 times that of computing  of the  method developed in the present paper, even when  $p$ is only $4$. We discussed  the relevant computational issue, the integral over all projection directions  in a test statistic in Wong, et al. (1995) and suggested  a number-theoretical method to reduce the computational workload. Xia (2009) also constructed a test that involved  searching for an optimal direction, but the test had no way of controlling  type I error.

Stute and Zhu (2002) considered a naive method to handle the curse of dimensionality when testing the parametric single-index model:  $Y=G({\beta}^{\top}X)+\epsilon$. Stemming from the  fact that under the null hypothesis, $
E(Y-G({\beta}^{\top}X) I(X\le t)=0$ for all $t \in R^p$ leads to $
E(Y-G({\beta}^{\top}X) I(\beta^{\top}X\le t)=0$ for all $t \in R^p$,
the test statistic is based on the  empirical process:
\begin{equation*}
R_{n}(x)= n^{-1/2}\sum_{i=1}^n(y_i- G(\hat{\beta}^{\top}x_i))I(\hat{\beta}^{\top}x_i \leq x),
\end{equation*}
where $\hat{\beta}$ is, under the null hypothesis, a root-$n$ consistent estimate of $\beta$. It has been proven to be powerful in many cases.
However, this test is a directional test rather than an omnibus test. Thus,  the general alternative of (\ref{(1.2)}) cannot be detected.
This phenomenon can be easily illustrated by the following alternative model:
$Y=\beta^{\top}_1X+ c\sin(\beta^{\top}_2X)+\epsilon$,
where $X$ is normally distributed  $N(0, I_p)$ with a $p\times p$ identity matrix $I_p$, and $\beta_1$ and $\beta_2$ are two orthogonal vectors. The value $c=0$ corresponds to the null hypothesis.
However, for any $c$, $E(Y-\beta^{\top}_1X|\beta^{\top}_1X)=0$. In other words, this conditional mean cannot distinguish between models under the null and alternative hypotheses.

However,  the advantage of SZ's test (Stute and Zhu, 2002) under the null hypothesis is very important particularly in high-dimensional paradigms as it fully uses the dimension reduction structure under the null.   Guo et al. (2015) recently proposed an adaptive-to-model  dimension-reduction test for the  model $Y=G({\beta}^{\top}X, \theta)+\epsilon$ against the general alternative model $Y=g(X)+\epsilon$.     The main idea is to fully utilize the dimension reduction structure about $X$ under the null hypothesis as Stute and Zhu (2002) did, but to adapt the alternative model such that the test is still  omnibus. Their test is based on a local smoothing technique. The improvement over existing local smoothing tests is significant. The test has a much faster convergence rate of $O(n^{-1/2}h^{-1/4})$  than the typical rate of $O(n^{-1/2}h^{-p/4})$ and can detect local alternatives distinct from the null hypothesis at the rate of $O(n^{-1/2}h^{-1/4})$ that is also much  faster than the typical rate of $O(n^{-1/2}h^{-p/4})$ that local smoothing tests can achieve. In other words, asymptotically, the test works as if  $X$ were univariate. Thus, the test  can  significantly avoid the curse of dimensionality. The numerical studies in their paper also indicated its advantages in cases with  moderate sample size. 

In the present paper, we consider a more general alternative model as
\begin{equation}\label{(1.3)}
Y = g(B^{\top}X, W) + \epsilon,
\end{equation}
where $B$ is a $p_1\times q$ matrix with $q$ orthogonal columns for an unknown number $q$ with $1\leq  q \leq p_1 $ and $g(\cdot)$ is still an unknown smooth function. To consider identifiability, assume that the matrix $B$ satisfies $B^{\top}B=I_{q}$. This model covers many popular models in the literature, such as the single-index models with $B=\beta$, the multi-index models with the absence of $W$, and partial single-index models with the mean function $g_1(\beta^TX)+g_2(W)$. $\beta$ is considered to be a column of $B$. When $q=p_1$ and $B=I_{p_1}$, the model~(\ref{(1.3)}) is reduced to the usual alternative model~(\ref{(1.2)}). The model~(\ref{(1.2)}) can then be rewritten as (\ref{(1.3)}). When $q=p_1$,
$g(X, W) = g(BB^{\top}X,W)  \equiv:  \tilde{g}(B^{\top}X,W),$
where $B$ is any $p_1\times p_1$ orthonormal  matrix. This persuasively demonstrates that the model~(\ref{(1.2)}) can  be treated as a special case of (\ref{(1.3)}). 
Based on this, a test can be constructed by noticing that under the null hypothesis, $E(Y-g(\beta^{\top}_1X, W,\theta)I(B^{\top}X\le t,W \le \omega)=0$ for all $(t,\omega)$ and under the alternative hypothesis, it is nonzero for some vector $(t, \omega)$.

To define an empirical version of this function as the basis for constructing a test statistic,  an adaptive estimate of $B$ is crucial for ensuring the test has the adaptive-to-model property. That is, we wish an estimate of $B$  to be consistent with $\kappa\beta$ for a constant $\kappa$ under the null and to $B$ under the alternative. Then,   under the null hypothesis,  the test can only rely on the dimension-reduced covariates  $(\beta^TX, W)$,  and  is still omnibus to detect the general alternative~(\ref{(1.3)}).   As mentioned above, when $W$ is absent, GWZ's test (Guo et al., 2014) has the adaptiveness property of the alternative model. To identify $B$ and its structural dimension,  various dimension reduction approaches such as  minimum average variance estimation (MAVE, Xia et al., 2002) and discretisation-expectation estimation (DEE, Zhu et al., 2010) have been suggested. However, when $W$ is present, these methods fail to work. Furthermore, due to the existence of $W$, even when the dimension $p_1=1$, the corresponding local smoothing test still has a slow convergence rate in the order of $O(n^{-1/2}h^{-(p_2+1)/4})$ where $p_2$ is the dimension of $W$.

In the present paper, we consider a global smoothing test that keeps the advantage of SZ's test, 
fully uses the dimension reduction structure and utilises an adaptive-to-model strategy to get the test omnibus. The key is to adaptively identify  $B$ such that under the null, $B$ is automatically identified to be $\beta$ to make the test dimension-reduced, and under the alternative, $B$ itself is identified to have the omnibus property.  To this end, the partial sufficient dimension reduction approach (Chiaromonte et al. (2002), Feng et al. (2013)) has to be applied.
To achieve the above target, we also need to identify or estimate the structural dimension $q$ of $B$. Under the null, $q=1$ is automatically identified or estimated. We then suggest a ridge-type eigenvalue ratio estimate. The details are presented in the next section. Another issue is critical value determination. In the present setting, the limiting null distribution is intractable, as it is for all global smoothing tests. A resampling approximation is required. We then propose a Monte Carlo approximation that also fully utilises the information in the hypothetical model so that the approximation can be as close to the sampling null distribution as possible.

The rest of the paper is organised as follows. In Section~2,  a dimension-reduction method, the partial discretization-expectation estimation,  is reviewed, and is then used to identify or estimate $B$. The   ridge-type eigenvalue ratio is also defined and its asymptotic properties are investigated in this section. Based on these, a test is constructed in Section~3.
The asymptotic properties under the null and local alternative hypotheses are also presented in this section. As the limiting null distribution is intractable, the Monte Carlo test approximation is described in Section~4. In Section~5, the simulation results are reported and a real data analysis is conducted for illustration. Technical proofs are found in the online supplementary material.

\section{Partial discretisation-expectation estimation and structural dimension estimation}
\subsection{A brief review on partial discretisation-expectation estimation }
As discussed above, identifying or estimating $B$ is important for  constructing an adaptive test. To this end, sufficient dimension reduction techniques can be applied. From the sufficient  dimension reduction theories, we can identify the space spanned by $B$, which is equivalent to,  $q$ basis vectors of the space spanned by $B$ (see, Chiaromonte et al., 2002).
Write $\tilde B$ as the $p\times q$  matrix consisting of these $q$ basis vectors. We call $\tilde B$ the  basis matrix. Note that $B$ is also a basis matrix of the space. Thus it is easy to see that for a $q\times q$ nonsingular matrix $C$, $\tilde B=B \times C^{\top}$. When $q=1$, $C$ is a constant and thus $\tilde B$ is a vector proportional to the vector $ \beta$ under the null. 
In Section~3 we show that identifying $\tilde B$ is enough for the testing problem described herein. In the following, $\tilde B$ is written  as $B$.

In this subsection, we focus on identifying a basis matrix $B$, this is equivalent to identifying the column space spanned by $B$.  This space is called the partial central subspace (first introduced by Chiaromonte et al., 2002), write as $\emph{S}^{(W)}_{Y|X}$. From their definition, it is the intersection of all subspaces $\emph{S}$ such that
$$Y \hDash X|(P_{\emph{S}}X,W),$$
where $\hDash$ stands for `independent of' and $P_{(\cdot)}$ indicates a projection operator with respect to the standard inner product. dim($\emph{S}^{(W)}_{Y|X}$) is called the structural dimension of  $\emph{S}^{(W)}_{Y|X}$. In our setup, the structural dimension is $1$ under the null and $q$ under the alternative.  Chiaromonte et al. (2002) and Wen and Cook (2007)  developed estimation methods for $\emph{S}^{(W)}_{Y|X}$ when $W$ is discrete. Li et al. (2010) proposed  groupwise dimension reduction (GDR), which can  also deal with this case. Feng et al. (2013) proposed partial discretisation-expectation estimation (PDEE) by extending discretisation-expectation estimation (DEE) in Zhu et al. (2010). All of those estimations use the root-$n$ consistency with  the partial central subspace. In this paper,  we adopt PDEE because PDEE is computationally inexpensive, and can be easily used to determine the structural dimension $q$. Also, when $W$ is absent, PDEE can naturally reduce to DEE without any changes in the algorithm.

From Feng et al. (2013), the following are the basic estimation steps.

\begin{enumerate}
\item Discretise the covariate $W= (W_1, \cdots, W_{p_2})$ into a set of binary variables by defining
 $W(\textbf{t})= (I\{W_1 \leq t_1\},\cdots, I\{W_{p_2} \leq t_{p_2}\})$ where the indicator functions $I\{W_i \leq t_i\}$ take  value 1 if $W_i \leq t_i$ and 0 otherwise, for $i=1,\cdots, p_2$.
\item Let $\emph{S}^{(W(t))}_{Y|X}$ denote the partial central subspace of $Y|(X, W(\textbf{t}))$, and $M(\textbf{t})$ be a $ p_1\times  p_1$ positive semi-definite matrix satisfying $\rm{Span}\{M(\textbf{t})\} =\emph{S}^{(W(\textbf{t}))}_{Y|X} $.
\item Let $T=\tilde W$ where $\tilde W$ is an independent copy of $W$. The target matrix is $M=E\{M(\tilde W)\}$. $ B$ consists of the eigenvectors that are associated with the nonzero eigenvalues of $M=E\{M(\tilde W)\}$.
\item Let ${w}_1,\cdots, {w}_{n}$ be the $n$ observations of $W$. Define an estimate of $M$ as
\begin{equation*}
M_{n}=\frac{1}{n}\sum^{n}_{i=1}M_n({w}_i),
\end{equation*}
where $M_n({w}_i)$ is the partial sliced inverse regression  matrix estimate defined in
Chiaromonte et al. (2002) where sliced inverse regression was proposed by Li(1991). Then when $q$ is given, an estimate $ B_n(q)$ of $B$ consists of the eigenvectors that are associated with the $q$ largest eigenvalues $\lambda_j$ of $M_{n}$.
$ B_n(q)$ can be root-$n$ consistent to $B$. For more details, readers may refer to Feng et al. (2013).
\end{enumerate}

\subsection{Structural dimension estimation}
The structural dimension $q$ is unknown in general. Interestingly,  even when it is given, we still want to estimate adaptively according to its values under the null and alternative because of its importance for the adaptive-to-model construction for the test. To estimate $q$, Feng et al. (2013) advised the BIC-type criterion that is an extension of that in Zhu et al (2006). However,  all practical uses show that selecting a proper penalty is not easy.
In this paper, we suggest a  ridge-type eigenvalue ratio estimate (RERE) to determine $q$ as:
\begin{eqnarray}\label{2.1}
\hat{q}=\arg\min_{1\leq j \leq p}\left\{ \frac{\hat{\lambda}^2_{j+1}+c_n}{\hat{\lambda}^2_j+c_n}\right\},
\end{eqnarray}
where $\hat{\lambda}_{p} \leq \cdots \leq \hat{\lambda}_{1}$ are the eigenvalues of the matrix $M_{n}$.
This method is motivated by Xia et al. (2015). The basic idea is as follows. Let $\lambda_j$ be the eigenvalues of the target matrix $M$. When $j\le q$, the eigenvalue $\lambda_j>0$ and thus, the ratio $r_{j-1}=\lambda_{j}/\lambda_{j-1}>0$; when $j> q$ $\lambda_j=0$. Therefore, $r_{q}=\lambda_{q+1}/\lambda_q=0$; and $\lambda_{j+1}/\lambda_{j}=0/0$. To define all ratios well, we can add a ridge in the ratio as $r_{j}=(\lambda_{j+1}+c_n)/(\lambda_{j}+c_n)$ for $1\le j\le p-1$. As $\hat \lambda_j^2$ converges to $\lambda_j^2$ at the rate of order $1/\sqrt n$ for $1\le j\le q$, and to $0$ at the rate of order $1/ n$ for $q+1\le j\le p$, then  $c_n=\log n/n$ can be a good choice. The algorithm is very easy to implement and the estimation consistency can be guaranteed.  The result is stated in the following. 

\begin{theorem}\label{Theorem1} Under Conditions A1 and A2 in the Appendix,
the estimate $\hat{q}$ of (\ref{2.1}) with $c_n=\log{n}/n$ has the following consistency:
\begin{itemize}
\item [(i)] under $H_0$,  $P(\hat{q}= 1)\rightarrow 1$;
\item [(ii)] under $H_1$,  $P(\hat{q}= q)\rightarrow 1 $.
\end{itemize}
\end{theorem}
From our justification presented in the Appendix,  the choice of $c_n$ can be in a relatively wide range to ensure consistency under the null and alternative hypotheses. However, to avoid the arbitrariness  of its choice, we find that $c_n=\log n /n$ is a proper choice.  The above identification of $q$ is very important for ensuring that the test statistic is adaptive to the underlying models.  
Finally, an estimate of $B$ is $B_n=B_n(\hat q).$ This estimate is used in the following test statistic construction.

\section{A partial dimension reduction adaptive-to-model test and its properties}

\subsection{Test statistic construction}
The hypotheses of interest can now be restated. The null hypothesis is
\begin{eqnarray*}
H_0:\ E(Y|X,W)= G(\beta^{\top}X,W, \theta)  \quad \rm for\ some \quad
\beta \in \mathbb{R}^{p_1},\ \theta \in \theta \in \mathbb{R}^d,
\end{eqnarray*}
against the alternative hypothesis: for any $\beta$ and $\theta$
\begin{eqnarray*}
H_1:\ E(Y|X,W)= g(B^{\top}X,W) \neq  G(\beta^{\top}X,W, \theta).
\end{eqnarray*}
In this subsection, let $\epsilon=Y-G(\beta^{\top}X,W, \theta)$ denote the error term under the null hypothesis.
Under $H_0$, $q=1$, and $B= \kappa\beta $ for some constant $\kappa$, then we have:
\begin{eqnarray*}
E(\epsilon|X,W)=0 &\Leftrightarrow& E(\epsilon |B^{\top}X,W)=0 \\
& \Leftrightarrow& E(\epsilon I\{(B^{\top}X,W) \leq (u,\omega)\})=0
\end{eqnarray*}
for all $(u,\omega)$.
Under $H_1$,  $E(Y - G(\beta^{\top}X,W,\theta)|X,W)=g(B^{\top}X,W)-G(\beta^{\top}X,W,\theta) \neq 0$, we then have:
\begin{equation*}
E(Y-G(\beta^{\top}X,W,\theta)|X,W) \neq 0 \Leftrightarrow E(Y-G(\beta^{\top}X,W,\theta)|B^{\top}X,W)\neq 0.
\end{equation*}

Before proceeding to the test statistic construction, recall that  what we can identify is $\tilde B=B\times C$ for a $q\times q$ orthogonal matrix $C$. Thus, we need to make sure this non-identifiability does not affect  the equivalence between  $E(Y-G(\beta^{\top}X,W,\theta)|\tilde B^{\top}X, W)\neq 0$ and $E(Y-G(\beta^{\top}X,W,\theta)|B^{\top}X, W) \neq 0$. This is easy to check. Note that  $\tilde B=B\times C^{\top}$ with $C$ being a non-singular matrix and thus $B$ and $\tilde B$ map  one-to-one. Then
\begin{eqnarray*}E(Y-G(\beta^{\top}X,W,\theta)|X, W)&=&E(g(B^{\top}X,W)-G(\beta^{\top}X,W,\theta)|X, W)\\
&=&E(\tilde g(\tilde B^{\top}X,W)-G(\beta^{\top}X,W,\theta)|X, W),
 \end{eqnarray*}
where $\tilde g(\cdot, \cdot)=g((C^{-1}\cdot, \cdot).$ It is equivalent between $E(Y-G(\beta^{\top}X,W,\theta)|B^{\top}X, W)\neq 0$ and $E(Y-G(\beta^{\top}X,W,\theta)|\tilde B^{\top}X, W)\neq 0$. Therefore, identifying $B$ itself is not necessary. As mentioned, we simply write $\tilde B$ as $B$.

Now we are in the position to define a residual-marked empirical process. Let
\begin{equation}\label{(3.1)}
V_{n}(u,\omega)=n^{-1/2}\sum_{i=1}^n(y_i- G(\beta^{\top}_{n}x_i,w_i,\theta_n))I\{(B_{n}(\hat q)^{\top}x_i,w_i) \leq (u,\omega)\},
\end{equation}
where $\beta_n$ and $\theta_n$ are the nonlinear least squares estimates respectively, and $B_n(\hat q)$ was defined before.

Therefore, we  use $V_{n}$ as the basis for constructing a test statistic:
\begin{eqnarray}\label{(3.2)}
T_n = \int{V^2_{n}(B_n(\hat q)^{\top}x,\omega)}dF_{ n}(B_n(\hat q)^{\top}x,\omega),
\end{eqnarray}
where $F_{ n}(\cdot)$ denotes the empirical distribution based on the samples $\{B_n(\hat q)^{\top}x_i, w_i\}_{i=1}^n$. Therefore, the null hypothesis is  rejected for large values of $T_n $.

It is clear that this test statistic is not scale-invariant and thus usually a normalizing constant is required. This constant needs to be estimated which involves many unknowns.  In this paper,  a Monte Carlo test procedure is recommended which can automatically make the test scale-invariant so that normalisation is not necessary. Additionally, it can mimic the sampling null distribution better than existing approximations such as that in Stute et al (1998).
The details can be found in  Section~4.

\subsection{Limiting null distribution}
To study the properties of the process $V_{n}(\cdot,\cdot)$ and the test statistic $T_n$, here we define a process for the purpose of theoretical investigation: for $u$ and $\omega$,
\begin{equation}\label{(3.3)}
V^0_{n}(u,\omega)= n^{-1/2}\sum_{i=1}^n(y_i- G(\beta^{\top}x_i,w_i,\theta))I\{(B^{\top}x_i,w_i) \leq (u,\omega)\}.
\end{equation}
When $E(Y^2) < \infty$, take the conditional variance of $Y$ given $B^{\top}X=u$ and $ W= \omega$,
$$\sigma^2(u,\omega) = Var(Y|B^{\top}X=u, W= \omega),$$
and put
$$ \psi (u,\omega) = \int^\omega_{-\infty}\int^u_{-\infty}{\sigma^2(v_1,v_2)}dF_{B^{\top}X,W}(v_1,v_2),$$
where $F_{B^{\top}X,W}(\cdot,\cdot)$ denotes the distribution function  of $(B^{\top}X,W)$.
It is easy to see that under $H_0$
$$\rm{Cov}\{V^0_{n}(u_1,\omega_1),V^0_{n}(u_2,\omega_2)\}= \psi (u_1 \wedge u_2,\omega_1 \wedge \omega_2).$$ By Theorem 1.1 in Stute (1997), we can assert that under $H_0$:
\begin{eqnarray}\label{(3.4)}
V^0_{n} \longrightarrow V_{\infty} \quad \rm in\ distribution, \quad
\end{eqnarray}
where $V_{\infty}$ is a continuous Gaussian process with mean zero and covariance kernel as follows: $$K((u_1,\omega_1),(u_2,\omega_2))=\psi (u_1 \wedge u_2, \omega_1 \wedge \omega_2).$$

\begin{theorem} \label{Theorem2}
 Under $H_0$ and the regularity conditions A1-A4 in the Appendix, we have the distribution
\begin{eqnarray*}
V_{n} \longrightarrow V_{\infty}-G^{\top}V \equiv V^1_{\infty},
\end{eqnarray*}
where $V_{\infty}$ is the  Gaussian process defined in (\ref{(3.4)}) and the vector-valued function $G^{\top}=(G_1,G_2,\cdots,G_{p+d})$ is defined as
\begin{eqnarray*}
G_i(u,\omega)=E\left[m_i(X,W, \beta, \theta)I\{(B^{\top}X,W) \leq (u,\omega)\}\right],
\end{eqnarray*}
where $B = \kappa\beta$
and $V$ is a $(p_{1}+d)-$dimensional normal vector with mean zero and covariance matrix $L(\beta, \theta)$ which is defined in the Appendix.
\end{theorem}
\begin{remark}
From this theorem, we can see that the test statistic has the same convergence rate of order $n^{-1/2}$ to its limit as that of existing global smoothing tests. In other words, in an asymptotic sense, there is no room for global smoothing tests to improve their convergence rate. Local and global smoothing tests differ in this feature, as $n^{-1/2}h^{p/4}$ can be much improved (Guo et al., 2015). However,  as in Stute and Zhu (2002), the new test can largely avoid the effect of  dimensionality to make the test more powerful when $p$ is large or even moderate. The simulations below illustrate this.
\end{remark}

\subsection{Power Study}
First, we present the asymptotic property under the global alternative hypothesis.
\begin{theorem} \label{Theorem3}
Under Conditions A1, A2, A3 and A4 and $H_{1n}$ with $C_n=c$ a fixed constant, we have in probability
\begin{eqnarray*}
n^{-1/2}V_{n}(u,\omega) \longrightarrow E[\{g(B^{\top}X,W)-G(\tilde{\beta}^{\top}X,W,\tilde{\theta})\}I\{(B^{\top}X,W) \leq (u,\omega)\}]
\end{eqnarray*}
where $(\tilde{\beta},\tilde{\theta})$ may be different from the true value $(\beta,\theta)$ under the null hypothesis. Then  $T_n\to \infty$ in probability.
\end{theorem}

To study how sensitive our new method is to the alternative hypotheses,  consider the following sequence of local alternatives:
\begin{eqnarray}\label{(3.5)}
H_{1n}:\ Y=G(\beta^{\top}X, W, \theta)+ C_n g(B^{\top}X, W)+\varepsilon,
\end{eqnarray}
where $C_n$ goes to zero.

Under the local alternatives with $C_n\to 0$, we also need to estimate  the structural dimension $q$. Recall that under the global alternative in Section~2, the estimate $\hat q=q$ had a probability going to zero, which could be larger than $1$ when $B$ contains more than one basis vector. However, under the above local alternatives, when $C_n$ goes to zero, the models converge to the hypothetical model that has one vector $\beta$. Thus, we anticipate  that $\hat q$  also converges to $1$  under the local alternatives.  The following lemma confirms this expectation.

\begin{lemma}\label{lemma1} Under  $H_{1n}$ in $(\ref{(3.5)})$, $C_n = n^{-1/2}$ and the regularity conditions in Theorem~\ref{Theorem2}, and  the estimate $\hat{q}$ in (\ref{2.1}) satisfies that as $ n\rightarrow \infty $, $P(\hat{q}= 1)\rightarrow 1$.
\end{lemma}

To further study the power performance of the test, assume an additional regularity Condition A5 in the Appendix.
\begin{theorem} \label{Theorem4}
 Under  $H_{1n}$ and Conditions A1, A2, A4 and A5, 
when $C_n=n^{-1/2}$, we have in distribution
\begin{eqnarray*}
V_{n}(u,\omega) \longrightarrow V_{\infty}(u,\omega) + E(g(B^{\top}X, W) I\{(\kappa \beta^{\top}X,W) \leq (u,\omega)\})+ G^{\top}(\eta-V)(u,\omega)
\end{eqnarray*}
where $V_{\infty}$, $G$ and $V$ are defined as those in Theorem~\ref{Theorem2}
and $\eta$ is a $(p_{1}+d)-$dimensional constant vector, which are defined in Appendix. Then $T_n$ has a finite limit.
\end{theorem}
\begin{remark}
This theorem shows that under the local alternatives, the test would also be directional, because  $\hat q$ is not a consistent estimate of $q$. This is caused by the difficulty of estimating $q$ when the alternative is too close to the null. If the estimation of $q$ could be improved, it is likely that  the omnibus property would still exist under the local alternative. We discuss this further in Section~6.
\end{remark}

\section{A Monte-Carlo test procedure}
As the limiting null distribution of the test statistic $T_n$ is not tractable, the nonparametric Monte Carlo test procedure is suggested to approximate the sampling null distribution, which is  similar in spirit to  the wild bootstrap, see  Stute et al. (1998) and  Zhu and Neuhaus (2000). However, to enhance the power of the test, we have a modified version that fully uses the model structure under the null.

A magical algorithm is developed to determine the $p-$values as follows:
\begin{itemize}
\item [{\it Step}] 1.
Generate a sequence of i.i.d variables $\textbf{U}=\{U_i\}^n_{i=1}$  from the standard normal distribution  $N(0,1)$. Then  construct the following process:
\begin{eqnarray*}
\Delta_n(u,\omega, \textbf{U}) = n^{-1/2} \sum^n_{i=1} \hat{\varrho}(x_i, w_i, y_i, \beta, \theta)U_i,
\end{eqnarray*}
where $\hat{\varrho}(x_i, w_i, y_i, \beta, \theta)$ is the estimate of $\varrho(x_i, w_i, y_i, \beta, \theta)$ and $\hat{\varrho}$ and $\varrho$ are defined as:
\begin{eqnarray*}
&&\varrho(x_i, w_i, y_i, \beta, \theta)= \epsilon_i I\{(B^{\top}_1x_i,w_i) \leq (u,\omega)\} - G^{\top}v_i,\\
&&\hat{\varrho}(x_i, w_i, y_i, \beta, \theta)= \hat{\epsilon}_i I\{(B^{\top}_{1n}x_i,w_i) \leq (u,\omega)\} - \hat{G}^{\top}\hat{v}_i,\\
&&G(u,\omega)=E\left[m(X,W, \beta, \theta)I\{(B^{\top}_1X,W) \leq (u,\omega)\}\right],\\
&&\hat{G}= n^{-1}\sum^n_{i=1}m(x_i,w_i, \beta_n, \theta_n)I\{(B^{\top}_{1n}X,W) \leq (u,\omega)\},\\
&&v_i=l(x_i,w_i, y_i,\beta, \theta),~~~~\hat{v}_i=l(x_i,w_i, y_i,\beta_n, \theta_n),\\
&&\hat{\epsilon}_i= y_i-G(\beta^{\top}_{n}x_i, w_i, \theta_n),,
\end{eqnarray*}
where $B_1$ and $B_{1n}$ denote the first column vectors of $B$ and $B_{n}(\hat q)$, respectively.
The resulting Monte Carlo test statistic is
$$\tilde{T}_n(\textbf{U}) = \int{\Delta^2_n(B^{\top}_{1n}x,w,\textbf{U})}dF_{B_{1n}}(x,w),$$
where $F_{B_{1n}}(\cdot)$ denotes the empirical distribution based on the samples $\{B^{\top}_{1n}x_i, w_i\}_{i=1}^n$.
\item [{\it Step}] 2. Generate $m$ sets of $\textbf{U}$,  $\textbf{U}_j$, $j=1,\cdots, m$, and get $m$ values of $\tilde{T}_n(\textbf{U})$, say $\tilde{T}_n(\textbf{U}_j)$, $j=1,\cdots, m$.
\item [{\it Step}] 3. The $p$-value is estimated by
\begin{eqnarray*}
 \hat{p} = m^{-1}\sum_{j=1}^mI(\tilde{T}_n(\textbf{U}_j)\geq T_n).
\end{eqnarray*}
Whenever $\hat{p} \leq \alpha $, reject $H_0$, for a given significance level $\alpha$, or the critical value is determined as the $(1-\alpha)100\%$ upper percentile of all $\textbf{U}_j$'s.
\end{itemize}
As mentioned before, this test procedure is scale-invariant although $T_n$ is not, because the resampling procedure does not need to  involve test statistic normalisation and $\hat{p} =  m^{-1}\sum_{j=1}^mI(\tilde{T}_n(\textbf{U}_j)\geq T_n)= m^{-1}\sum_{j=1}^mI(\tilde{T}_n(\textbf{U}_j)/c\geq T_n/c)$ for any $c>0$.
\begin{remark}
It is worth pointing out that the algorithm is different from   traditional nonparametric Monte Carlo test procedures that use the vector $B(\hat q)_{n}^{\top}X$. More details can be found in  Zhu (2005). When we only use the vector $B_{1n}$, which is associated with the largest eigenvalue of the target matrix $M_n$ defined in Section~2, we only use univariate $B_{1n}^{\top}X$, which is $\beta^{\top}X$ under the null asymptotically. This makes the approximation as close to the sampling null distribution as possible.
\end{remark}

The following theorem states the consistency of the conditional distribution approximation even under local alternatives.
\begin{theorem} \label{Theorem3}
Under the conditions in Theorem~\ref{Theorem2} and the null hypothesis or the local alternative hypothesis with $C_n=n^{-1/2}$, we know that for almost all sequences
 $\{(y_1, x_1, w_1), \cdots , (y_n, x_n, w_n), \cdots\},$
the conditional distribution of $\tilde{T}_n(\textbf{U})$ converges to the limiting null distribution of $T_n$.
\end{theorem}
\underline{}\section{Numerical Studies}
\subsection{Simulations}
In this subsection, we conduct simulations to examine the finite-sample performance of the proposed test. The simulations are based on $2000$ Monte Carlo test replications to compute the critical values or $p$ values. Each experiment is then repeated $1000$ times to compute the empirical sizes and powers at the significance level $\alpha = 0.05$. To estimate the central subspace spanned by $B$, we use the SIR-based PDEE/DEE procedure according to the cases with and without the variate $W$ in the model. In these two cases, we call the test $T^{PDEE}_{n}$.

We choose ZH's test (Zheng, 1996) and SZ's test ( Stute and Zhu, 2002) as the representatives of local and and global smoothing tests, respectively, to compare with our test. We choose these tests because 1). ZH's test has the explicitly and tractable limiting null distribution that can be used  to determine  the critical values; 2). like other local smoothing tests, the re-sampling version helps improve its performance (we then also include the re-sampling version of ZH's test); and 3) SZ's test is asymptotically distribution-free and powerful in many situations, but is not an omnibus test. We also compare our test to GWZ's test (Guo et al., 2015), because it is based on ZH's test but also has the adaptive-to-model property, it can be much more powerful. We write the proposed test, ZH's, SZ's and GWZ's tests as  $T^{PDEE}_{n}$, $T^{ZH}_{n}$, $T^{SZ}_{n}$ and $T^{GWZ}_{n}$, respectively.

In this section, we first design four examples to examine the performance in four scenarios without the random variable $W$. The first example has the same projection direction in both the hypothetical and alternative model. The second example is used to check the adaptiveness of our test to omnibus testing even when  dimension reduction structure under the null is fully adopted, showing that SZ's test is directional and thus has much less power. The third example is used to check the effect of dimensionality from $X$ for local smoothing tests, and to compared against with  ZH's and GWZ's tests. The fourth example is used to assess the effect of correlations among the components of $X$. In the first three examples, the data $(x_i,w_i)$ are generated from the multivariate standard normal distribution $N(0, I_p)$, independent of the standard normal errors $\epsilon_i$.

{\it Example} 1.  Consider the following regression model:
 \begin{itemize}
 \item  $Y= \beta^{\top}_0X + a \times cos(0.6 \pi \beta^{\top}_0X) + 0.5\times \epsilon$ and $\beta_0 = (0, 0,1,1)/\sqrt{2}$.
 \end{itemize}
 The values $a=0, 0.2,0.4,0.6,0.8, 1$ are used. The value $ a  = 0$ corresponds to the null hypothesis and  $a\neq 0$ to the alternative hypothesis.
The power function is plotted in  Figure (\ref{figure1}).
\begin{center}
Figure~(\ref{figure1}) about here
\end{center}
Some findings are as follows. The  power increases reasonably with larger $a$. The proposed test $T^{PDEE}_{n}$ is significantly and uniformly more powerful  than  $T^{ZH}_{n}$ and $T^{SZ}_{n}$. When $a$ is not large, $T^{SZ}_{n}$ works better than $T^{ZH}_{n}$, and when $a$ is large, $T^{ZH}_{n}$ slightly outperforms $T^{SZ}_{n}$ in power.

{\it Example} 2. To further check the omnibus property of the proposed test to detect general alternative models,  a comparison with  SZ's test and ZH's test is again carried out. In this example,  we generate the data from the following regression model:
\begin{itemize}
\item  $Y= \beta^{\top}_0X + a \times 0.125\exp(0.3\beta^{\top}_1X) + 0.5\times \epsilon;$
\end{itemize}
where $\beta_0 = (1, 1,0,0)/\sqrt{2}$ and $\beta_1 = (0, 0, 1, 1)/\sqrt{2}$. The values $a=0,0.1, 0.2,0.3,0.4,$ $0.5,0.6,0.7,0.8,0.9, 1$ are used.
 In this model, $B=(\beta^{\top}_0, \beta^{\top}_1)^{\top}$ and $\beta^{\top}_0X$ is orthogonal to the functions under the alternatives. We can see that SZ's test cannot detect such alternatives.  The results are reported in Figure~\ref{figure2}.
\begin{center}
Figure~(\ref{figure2})   about here
\end{center}
The results clearly show  that SZ's test $T^{SZ}_n$ and ZH's test $T^{SZ}_n$ are not very sensitive to the alternatives. In particular, when the sample size is small ($n=100$),  SZ's test $T^{SZ}_n$  has almost no power.

\newpage
{\it Example} 3. To gain further insights into our test, we consider  the effect of the dimensionality of $X$. When the number of dimensions is large, ZH's test does not maintain the significance level or power performance, due to slow convergence. Thus,  the wild bootstrap is applied to approximate the sampling null distribution. The re-sampling time is $2000$ in this simulation study. The bootstrap version is written as $T^{ZHB}_{n}$. 
GWZ's test  is also compared. 

 Consider the  models:
\begin{itemize}
\item  $Y= \beta^{\top}_0X + a \times\{0.3(\beta^{\top}_1X)^3+0.3(\beta^{\top}_1X)^2\} + 0.5\times \epsilon;$
\end{itemize}
where $\beta_0 = (1, 1,1, 1,0,0,0,0)/2$ and $\beta_1 = (0,0,0,0, 1, 1,1, 1)/2$. Then the dimension $p=8$. The results are listed in Table~\ref{table1}.
\begin{center}
Table~(\ref{table1}) about here
\end{center}
From Table~\ref{table1}, we can see that $T^{ZH}_{n}$ does not maintain the significance level well, but its bootstrap version $T^{ZHB}_{n}$ and $T^{GWZ}_{n}$ can, and $T^{PDEE}_{n}$ works better uniformly. {
The dimension reduction adaptive-to-model test $T^{GWZ}_{n}$ has a clear advantage over its counterpart $T^{ZH}_{n}$ in  maintaining the significance level and gaining power. However, $T^{PDEE}_{n}$ still works better uniformly. This seems to suggest that the global smoothing test performs better than the local smoothing test when both are constructed via the dimension reduction technique.  Compared with the results in Figures~\ref{figure1} and \ref{figure2} with $p=4$,  the dimension $p$ has little effect for $T^{PDEE}_n$. However, it has a very significant effect for $T^{ZHB}_{n}$ and $T^{ZH}_{n}$.  When the number of dimensions is higher, the performance of $T^{ZH}_{n}$ and $T^{ZHB}_{n}$ is worse.
We do not include the simulation results to save space. 

{\it Example} 4. To further assess the performance of the test $T^{PDEE}_{n}$, we consider the
effect of the correlated covariate $X$ and the  distribution of the error term $\epsilon$. Consider the following model:

\begin{itemize}
\item $y=\beta_0^{\top}X + a\times\exp(-(\beta_0^{\top}X)^2/2)/2 + 0.5\times\epsilon;$
\end{itemize}
where $X$ follows a normal distribution $N(0, \Sigma)$ with the covariance matrix $\Sigma_{ij}=I(i=j)+ \rho^{|i-j|} I(i \neq j)$ for $\rho=0.5$, $i,j =1, 2,\cdots, p$, $\beta_0=(1, 1, -1, -1)/2$ and $\epsilon$ follows the student's t-distribution with 4 degrees of freedom.
\begin{center}
Table~(\ref{table2}) about here
\end{center}
The results are presented in Table~\ref{table2}. Comparing the results in this table with those in Figures~\ref{figure1} and \ref{figure2}, it is clear that with the correlated covariate $X$, we arrive at similar conclusions to those in {\it Examples}~1 and 2. $T^{PDEE}_{n}$ easily maintains the significance level. {
We also find that when the structural dimension $q=1$ under the alternative hypothesis, the  power performance of $T^{GWZ}_{n}$ is very similar to that of $T^{PDEE}_{n}$. Comparing {\it Example}~3 in  Table~\ref{table1}  with {\it Example}~4 in  Table~\ref{table2}, we can see that the lower structural dimension increases  $T^{GWZ}_{n}$'s  empirical power. This suggests that the structural dimension $q$ still has a negative effect on $T^{GWZ}_{n}$, although theoretically, $T^{GWZ}_{n}$ can detect alternatives distinct from the null at the same rate as if the dimension of $X$ were one. However, the power of $T^{PDEE}_n$ does not deteriorate when the  the structural dimension is increased. 
Further, $T^{PDEE}_n$ can control type I error very well and is significantly more powerful than ZH's  and SZ's tests. It is evident that $T^{PDEE}_{n}$ is robust to the error term.

In summary, the global smoothing-based dimension reduction adaptive-to-model test inherits the advantages of global smoothing tests and has the adaptive-to-model property when the dimension reduction structure is adopted.

Now we consider the parallel models in {\it Examples}~1-4 when the covariate $W$ is included. However, we present only the results for  $T^{PDEE}_n$ because based on the  results in the above examples and comparisons, the performance of the competitors is even worse when  there are $q_1$ more dimensions in the model (meaning that $q_1$ more dimensions are added when $W$ is $q_1$-dimensional).

{\it Example} 5. The four models are:
\begin{itemize}
 \item []Case 1). $Y= \beta^{\top}_0X + W + a \times cos(0.6 \pi \beta^{\top}_0X) + 0.5\times \epsilon$;
\item []Case 2). $Y= \beta^{\top}_0X + \sin(W) + a\times (0.5(\beta^{\top}_1X)^2 + 2\sin(W))+0.5\times \epsilon$;
\item []Case 3). $Y= \beta^{\top}_0X + \cos(W) + a \times\{0.3(\beta^{\top}_1X)^3+0.3(\beta^{\top}_1X)^2\} + 0.5\times \epsilon$;
\item []Case 4). $y=\beta_0^{\top}X + \sin(W) + a\times\exp(-(\beta_0^{\top}X)^2/2)\times W + 0.5\epsilon$.
\end{itemize}
All of the settings are the same as the respective settings in {\it Examples} 1-4 except for the additional $W$ following the normal distribution $N(0,1)$.  The results are reported in Table~\ref{table3}.

\begin{center}
Tables~(\ref{table3}) about here
\end{center}

The reported results clearly indicate that when $W$ is presented, $T^{PDEE}_n$ still works well in  maintaining the significance level and detecting general alternatives.
\subsection{Real Data Analysis}
In this subsection, for illustration  we perform the regression modelling of the well-known Boston Housing Data, initially studied by Harrison and Rubinfeld (1978)..
The data set contains 506 observations and 14 variables, as follows:  the median value of owner-occupied homes in \$1000's ({\it MEDV}),  per capita crime rate by town ({\it CRIM}), proportion of residential land zoned for lots over 25,000 sq.ft. ({\it ZN}),proportion of non-retail business acres per town ({\it INDUS}), Charles River dummy variable (1 if tract bounds river; 0 otherwise) ({\it CHAS}), nitric oxides concentration (parts per 10 million) ({\it NOX}),  average number of rooms per dwelling ({\it RM}),  proportion of owner-occupied units built prior to 1940 ({\it AGE}),  weighted distances to five Boston employment centres ({\it DIS}), index of accessibility to radial highways ({\it RAD}), full-value property-tax rate per $10,000$ ({\it TAX}),  pupil-teacher ratio by town ({\it PTRATIO}), the proportion of black people by town ({\it B}) and lower status of the population ({\it LSTAT}).

As suggested by Feng et al. (2013),  we take the logarithm of ({\it MEDV}) as the predictor, the predictor {\it CRIM} as $W$ and the other 11 predictors as $X$, except {\it CHAS}, because  it has little influence on  the housing price as advised  by Wang et al. (2010), and is thus excluded from this data analysis. In this data analysis, we standardise the predictors for ease of explanation. From the plot in Feng et al (2013), a simple linear model is considered to be the hypothetical model. The SIR-based PDEE procedure is applied to determine the partial central subspace $\emph{S}^{(W)}_{Y|X}$. The structural dimension $\hat{q}=2$ of the partial central subspace is determined by  RERE in Section~2. A total of $2000$ Monte Carlo test replications are implemented to compute the $p$ value, which is about zero. Hence, it is reasonable to reject the null hypothesis. Moreover, $\hat{q}$ is estimated to be 2. Thus, partial multi-index modelling is required although the plot seems to suggest a linear model.

\section{Discussions}

In this paper, we propose an adaptive-to-model dimension reduction test based on a residual marked empirical process for partially parametric single-index models. The test can fully utilise the dimension reduction structure to reduce dimensionality problems, while remaining an omnibus test.
Comparisons with existing local and global smoothing tests suggest that 1). model-adaptation enhances the power performance, also maintaining the significance level; and 2). the global smoothing-based adaptive-to-model test outperforms the local smoothing-based adaptive-to-model test. Thus, a global smoothing test is worthy of recommendation.
 This method can be readily applied to other models and problems when a dimension reduction structure is presented. The research is on-going.

In the hypothetical and alternative model, the independence between the error and the covariates is assumed. This condition is fairly strong.  The condition can be weakened to handle the testing problem for the following hypothetical and alternative models:
\begin{eqnarray*}
&&Y = G(\beta^{\top}X, W, \theta) + \delta(\beta^{\top}X, W)\epsilon,\\
&&Y = g(B^{\top}X, W) + \delta(B^{\top}X, W)\epsilon.
\end{eqnarray*}
Here, all of the settings are the same as those considered in the present paper,
except that the function $\delta(\cdot)$ is an unknown smooth function. $ B_n(\hat{q})$, estimated by the SIR-based PDEE/DEE procedure is still a root-$n$ consistent estimate of $B$. Thus, the  proposed  test can still be feasible.

Further, we still find a theoretical shortcoming in the omnibus property of the proposed test as discussed before. That is, under the local alternatives that converge to the null at a certain rate, the proposed test, unlike existing omnibus tests, cannot be powerful,  because under the local alternatives with $C_n= 1/\sqrt{n}$,  the method can only estimate $q$ to be $1$. Thus, the estimate $\hat B$ converges to $\beta$, and when the other directions in $B$ are orthogonal to $\beta$ and the function has some special structure, our test may not have good power. However, this does not mean that our test cannot detect any local alternative models. When the convergence rate $C_n$ becomes slower, the RERE can still estimate $B$ well by choosing a suitable ridge value $c_n$ and then the alternatives can be detected.  Research is ongoing to derive a more powerful test under local alternatives.

\

\newpage

\leftline{\large\bf References}

\begin{description}

\item Bierens, H. J. (1990). A consistent conditional moment test of functional form. {\it Econometrica}. {\bf 58}, 1443-1458.

\item Chiaromonte, F., Cook, R. D. and Li, B. (2002). Sufficient dimension reduction in
regressions with categorical predictors. {\it The Annals of Statistics}. {\bf 30}, 475-497.



\item Dette, H. (1999). A consistent test for the functional form of a regression based on a difference of variance estimates. {\it The Annals of Statistics}. {\bf 27}, 1012-1050.


\item Escanciano, J. C. (2006). A Consistent Diagnostic Test for Regression Models
Using Projections. {\it Econometric Theory}. {\bf 22}, 1030-1051.

\item Escanciano, J. C. (2007). Model Checks Using Residual Marked Empirical Processes. {\it Statistica Sinica}. {\bf 17}, 115-138.

\item Fan, J. Q. and Huang, L. S. (2001). Goodness-of-fit tests for parametric regression
models, {\it Journal of the American Statistical Association}, {\bf 96}, 640-652.

\item Fan, Y. and  Li, Q., (1996). Consistent model specication tests: omitted variables and semiparametric functional forms. {\it Econometrica}, {\bf 64}, 865-890.

 \item Fan, J., Zhang, C. and Zhang, J. (2001) Generalized likelihood ratio
statistics and Wilks phenomenon. {\it The Annals of Statistics}, {\bf 29}, 153-193.

\item Feng, Z., Wen, X., Yu Z. and Zhu, L. X. (2013). On Partial Sufficient Dimension
Reduction With Applications to Partially Linear Multi-Index Models. {\it Journal of the
American Statistical Association}. {\bf 501}, 237-246.

\item Guo, X., Wang, T. and Zhu, L. X. (2015). Model checking for generalized linear models: a
dimension-reduction model-adaptive approach. \url{http://www.math.hkbu.edu.hk/~lzhu/}

\item Gonz\'{a}lez-Manteiga, W. and Crujeiras, R. M. (2013). An updated review
of Goodness-of-Fit tests for regression models. {\it TEST}, {\bf 22}, 361-411.

\item H\"{a}rdle, W. and  Mammen, E. (1993). Comparing nonparametric versus
parametric regression fits. {\it The Annals of Statistics}, {\bf 21}, 1926-1947.

\item Harrison, D. and Rubinfeld, D. L. (1978). Hedonic prices and the demand for clean
air. {\it Journal of Environmental Economics and Management}, {\bf 5}, 81-102.

\item Huber, P. J. (1985). Projection pursuit. {\it The Annals of Statistics}, 435-475.

\item Khmaladze, E. V. and Koul, H. L. (2004). Martingale transforms goodness-of-fit tests in regression models. {\it The Annals of Statistics}, {\bf 37}, 995-1034

\item Koul, H. L. and Ni, P. P. (2004). Minimum distance regression model checking. {\it Journal of Statistical Planning and Inference}, {\bf 119}, 109-141.


\item Li, L., Li, B. and Zhu, L. X. (2010). Groupwise dimension reduction. {\it Journal of the
American Statistical Association}. {\bf 105}, 1188-1201.


\item Lavergne, Q., and Patilea, V. (2008). Breaking the Curse of Dimensionality in
Nonparametric Testing. {\it Journal of Econometrics}. {\bf 143}, 103鈥?22.

\item Lavergne, P. and Patiliea, V. (2012), One for all and all for one: Regression checks with many regressors. {\it Journal of business \& economic statistics}, {\bf 30}, 41-52.






\item Stute, W. (1997). Nonparametric model checks for regression. {\it The Annals of Statistics}. {\bf 25}, 613-641.

\item Stute, W., Gonz\'{a}les-Manteiga, W. and Presedo-Quindimil, M. (1998). Bootstrap approximation in model checks for regression. {\it Journal of the American Statistical Association}. {\bf 93}, 141-149.


\item Stute, W. and Zhu, L. X. (2002). Model checks for generalized linear models. {\it Scandinavian Journal of Statistics }. {\bf 29}, 535-545.

\item Stute, W. and Zhu, L. X. (2005). Nonparametric checks for single-index models,
{ \it The Annals of Statistics}, {\bf 33}, 1048-1083.

\item Van Keilegom, I., Gonz\'{a}les-Manteiga, W. and S\'{a}nchez Sellero, C. (2008). Goodness-of-fit tests in parametric regression based on the estimation of the error distribution. {\it TEST}, {\bf 17}, 401-415.

\item  Wang, J. L., Xue, L. G., Zhu, L. X. and Chong, Y. S. (2010). Estimation for a partial linear
single-index model. {\it  The Annals of Statistics}, {\bf 30}, 475-497.

\item Wong, H. L., Fang, K. T. and Zhu, L. X. (1995). A test for multivariate normality based on sample entropy and projection pursuit. {\it Journal of Statistical Planning and Inference}, {\bf 45}, 373 - 385.
\item Wen, X. and Cook, R. D. (2007). Optimal sufficient dimension reduction in regressions
with categorical predictors. {\it Journal of Statistical Planning and Inference}. {\bf 137},
1961-1978.


\item Xia, Q., Xu, W. L. and Zhu. L. X (2015). Consistently determining the number of factors in multivariate volatility modelling, {\it Statistica Sinica}, {\bf 25},1025-1044.

\item Xia, Y. C.(2009). Model check for multiple regressions via dimension reduction.  {\it Biometrika}, {\bf 96}, 133-148.

\item Xia, Y. C., Tong, H., Li, W. K. and Zhu, L. X. (2002). An adaptive estimation of dimension reduction space. {\it Journal of the Royal Statistical Society: Series B}, {\bf 64}, 363-410.



\item Zheng, J. X. (1996). A consistent test of functional form via nonparametric
estimation techniques. {\it Journal of Econometrics}, {\bf 75}, 263-289.

\item Zhu, L. P., Zhu, L.X. , Ferr\'{e}, L. and Wang, T. (2010). Sufficient dimension reduction through discretization-expectation estimation. {\it Biometrika}, {\bf 97}, 295-304.

\item Zhu, L. X. (2005). Nonparametric Monte Carlo tests and their applications.
{\it Springer, New York}.

\item Zhu, L. X. and An, H. Z. (1992). A test  for nonlinearity in regression models. {\it Journal of Mathematics}, {\bf 4}, 391-397. (Chinese).


\item Zhu, L. X. and Li, R. Z. (1998). Dimension-reduction type test for linearity
of a stochastic model. {\it  Acta Mathematicae Applicatae Sinica.} {\bf 14}, 165-175.

\item Zhu, L. X. and Neuhaus, G. (2000). Nonparametric Monte Carlo tests for multivariate distributions.
    {\it  Biometrika}, {\bf 87}, 919-928.


\end{description}

\
\newpage

\begin{figure}
  \centering
  \includegraphics[width=\textwidth]{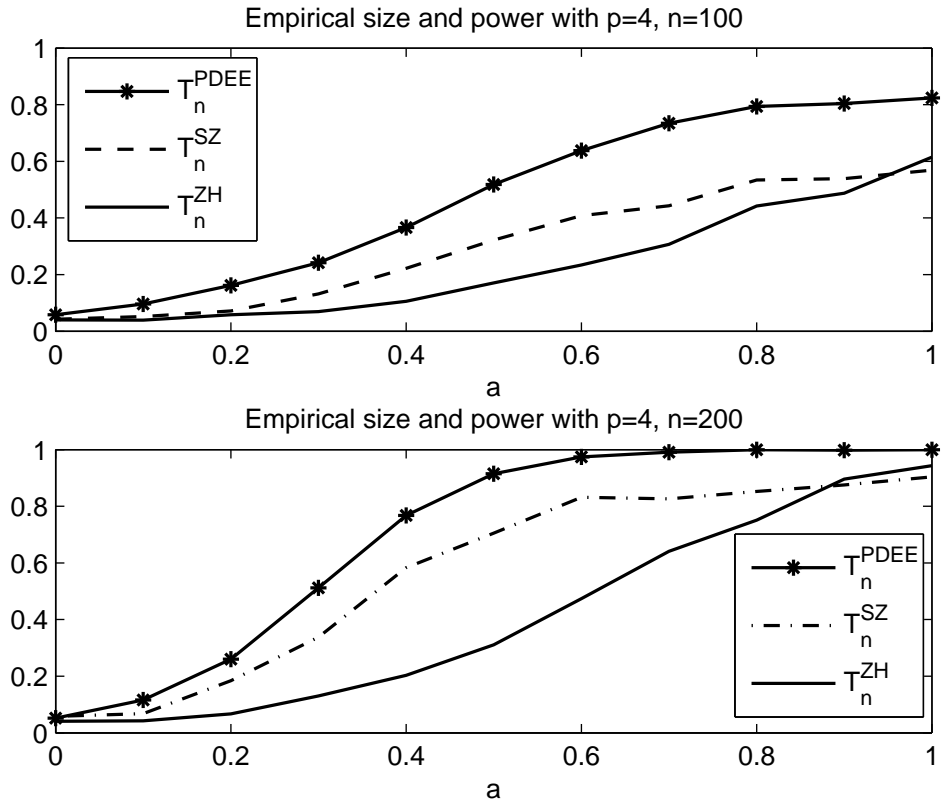}\\
  \caption{The empirical size and power curves of $T^{PDEE}_n$, $T^{SZ}_n$ and $T^{ZH}_n$ in Example~1. 
  }\label{figure1}
\end{figure}

\begin{figure}
  \centering
  \includegraphics[width=13.5cm]{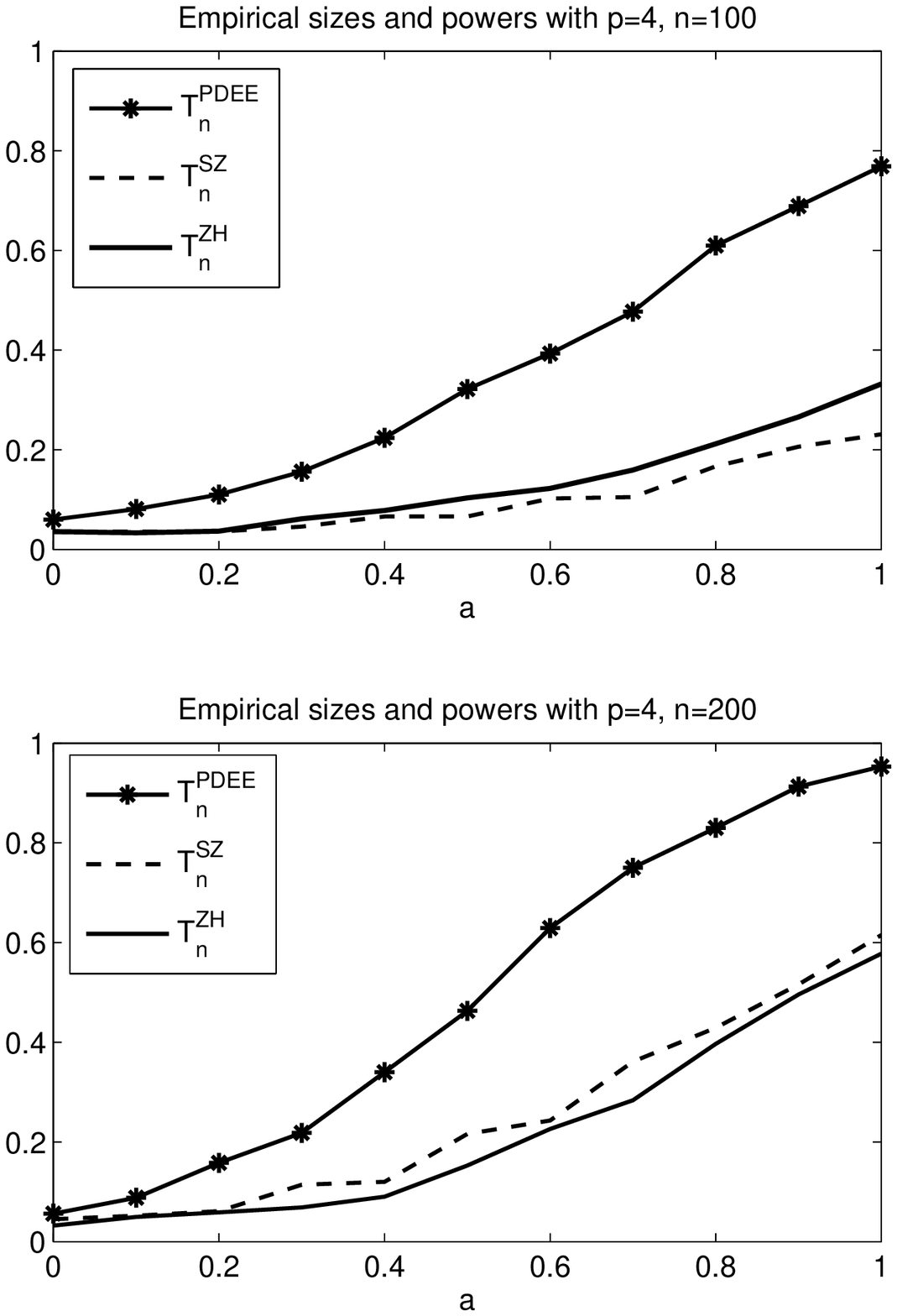}\\
  \caption{The empirical size and power curves of $T^{PDEE}_n$, $T^{SZ}_n$ and $T^{ZH}_n$ in Example~2. }\label{figure2}
\end{figure}

\begin{table}[htb!]\caption{ Empirical sizes and powers of $T^{PDEE}_{n}$, $T^{ZHB}_{n}$, $T^{ZH}_{n}$ and $T^{GWZ}_{n}$ for  Example~3 with $p=8$. \label{table1}
\vspace{0.35cm}}
\centering
 {\tiny\scriptsize\hspace{12.5cm}
\renewcommand{\arraystretch}{1}\tabcolsep 0.2cm
\begin{tabular}{cccccccccccccc}
\hline
&\multicolumn{1}{c}{$a$}&\multicolumn{3}{c}{$T^{PDEE}_{n}$} & \multicolumn{3}{c}{$T^{ZHB}_{n}$} &
   \multicolumn{3}{c}{$T^{ZH}_{n}$}&\multicolumn{3}{c}{$T^{GWZ}_{n}$}\\
$n$&&50 &100& 200& 50 &100& 200&50 &100& 200&50 &100& 200 \\
\hline
&0  &0.0560&0.0570&0.0460&0.0400&0.0450&0.0580&0.0240&0.0270&0.0430&0.0350&0.0610&0.0510 \\
&0.2&0.2410&0.2530&0.3820&0.0480&0.0630&0.0810&0.0260&0.0390&0.0600&0.0760&0.1100&0.1830 \\
&0.4&0.4220&0.5660&0.8460&0.0650&0.1140&0.1700&0.0460&0.0810&0.1430&0.1220&0.2750&0.4710 \\
&0.6&0.5450&0.7460&0.9650&0.0980&0.1810&0.3860&0.0690&0.1670&0.3590&0.2120&0.4460&0.7680 \\
&0.8&0.6360&0.8750&0.9870&0.1230&0.2670&0.5380&0.1070&0.2730&0.5780&0.2840&0.6070&0.9230 \\
&1.0&0.7230&0.9290&0.9930&0.1600&0.3510&0.7220&0.1330&0.3820&0.7330&0.3920&0.7440&0.9720 \\
 \hline
\end{tabular}
}
\end{table}

\begin{table}[htb!]\caption{Empirical sizes and powers of $T^{PDEE}_{n}$, $T^{SZ}_{n}$, $T^{ZH}_{n}$ and $T^{GWZ}_{n}$ for  Example~4 with $p=4$ and correlated covariates. \label{table2}
\vspace{0.35cm}}
\centering
 {\tiny\scriptsize\hspace{12.5cm}
\renewcommand{\arraystretch}{1}\tabcolsep 0.2cm
\begin{tabular}{cccccccccccccc}
\hline
&\multicolumn{1}{c}{$a$}&\multicolumn{3}{c}{$T^{PDEE}_{n}$} & \multicolumn{3}{c}{$T^{SZ}_{n}$} &
   \multicolumn{3}{c}{$T^{ZH}_{n}$}&\multicolumn{3}{c}{$T^{GWZ}_{n}$}\\
\hline
$n$&&50 &100& 200& 50 &100& 200&50 &100& 200 &50 &100& 200\\
\hline
&0  &0.0620&0.0570&0.0520&0.0310&0.0390&0.0500&0.0380&0.0390&0.0410&0.0510&0.0530&0.0450\\
&0.2&0.1020&0.1670&0.2070&0.0520&0.0600&0.1390&0.0550&0.0690&0.0820&0.0810&0.1110&0.1670\\
&0.4&0.2350&0.4160&0.5870&0.0920&0.1560&0.4060&0.0880&0.1490&0.2720&0.1530&0.2980&0.5220\\
&0.6&0.4310&0.6600&0.8850&0.1560&0.3780&0.7260&0.2150&0.3740&0.5950&0.2980&0.6110&0.8520\\
&0.8&0.5820&0.8540&0.9780&0.2680&0.5420&0.9180&0.3690&0.5860&0.8850&0.5410&0.8280&0.9690\\
&1.0&0.6960&0.9510&0.9960&0.2920&0.7300&0.9800&0.5300&0.7830&0.9640&0.7040&0.9610&0.9990\\
 \hline
\end{tabular}
}
\end{table}

\begin{table}[htb!]\caption{Sizes and powers of $T^{PDEE}_{n}$ for Example~5. \label{table3}
\vspace{0.75cm}}
\centering
 {\small\scriptsize\hspace{12.5cm}
\renewcommand{\arraystretch}{1}\tabcolsep 0.5cm
\begin{tabular}{ccccccc}
\hline
&\multicolumn{1}{c}{$a$}&\multicolumn{1}{c}{n=50} &\multicolumn{1}{c}{n=100}  &
  \multicolumn{1}{c}{n=200} &
  \multicolumn{1}{c}{n=400} \\
\hline
Case 1 with $p=4$
& 0     &0.0680    &0.0590    &0.0560 &0.0500\\
and $q=1$ & 0.2   &0.1720    &0.1720    &0.2940 &0.6420\\
& 0.4   &0.2620    &0.5120    &0.8740 &1.0000\\
& 0.6   &0.5080    &0.8950    &1.0000 &1.0000\\
& 0.8   &0.6520    &0.9640    &1.0000 &1.0000\\
& 1     &0.7020    &0.9840    &1.0000 &1.0000\\
\hline
Case 2 with $p=4$

&0     &0.0580  &0.0550    &0.0530    &0.0480 \\
and $q=2$ &0.2   &0.0690  &0.0960    &0.1390    &0.3760 \\
&0.4   &0.1100  &0.2460    &0.6910    &0.9970 \\
&0.6   &0.2120  &0.5730    &0.9850    &1.0000 \\
&0.8   &0.3040  &0.8660    &1.0000    &1.0000 \\
&1     &0.4210  &0.9380    &1.0000    &1.0000 \\

\hline
Case 3 with $p=8$
&0    &0.0630  &0.0560    &0.0550    &0.0500 \\
and $q=2$ &0.2  &0.1330  &0.1820    &0.2960    &0.5500 \\
&0.4  &0.2450  &0.4150    &0.7190    &0.9520 \\
&0.6  &0.3570  &0.5970    &0.9110    &0.9980 \\
&0.8  &0.4210  &0.7240    &0.9510    &1.0000 \\
&1    &0.5010  &0.8020    &0.9780    &1.0000 \\

\hline
Case 4 with $p=4$
&0   &0.0600  &0.0450    &0.0470    &0.0510\\
and $q=1$ &0.2 &0.1020  &0.1540    &0.2700    &0.4960\\
&0.4 &0.2170  &0.4270    &0.7440    &0.9820\\
&0.6 &0.4050  &0.7240    &0.9730    &1.0000\\
&0.8 &0.5410  &0.8970    &0.9990    &1.0000\\
&1.0 &0.6460  &0.9560    &0.9990    &1.0000\\
\hline

\end{tabular}
}
\end{table}

\end{document}